\documentclass[aps,prd,superscriptaddress,showpacs, preprint]{revtex4-1}
\usepackage{graphicx}
\usepackage{epstopdf}
\usepackage{amsmath}
\usepackage{amsfonts} 
\usepackage{amssymb}
\usepackage{latexsym}
\usepackage{hyperref}
\usepackage[english]{babel}
\usepackage[utf8]{inputenc}
\usepackage{xcolor}
\usepackage{slashed}
\usepackage{bm}

\begin{document}

\title{Effective theories and non-minimal couplings  in  low-dimensional systems}

\author{M.G. Campos}\email{mgcampos@cbpf.br}
\affiliation{Centro Brasileiro de Pesquisas F\'{i}sicas (CBPF), Rua Dr. Xavier Sigaud 150, Urca, Rio de Janeiro, Brazil, CEP 22290-180}

\author{L.P.R. Ospedal}\email{leoopr@cbpf.br}
\affiliation{Centro Brasileiro de Pesquisas F\'{i}sicas (CBPF), Rua Dr. Xavier Sigaud 150, Urca, Rio de Janeiro, Brazil, CEP 22290-180}

\begin{abstract}

Dimensionality aspects of non-minimal electromagnetic couplings are investigated. 
By means of the Foldy-Wouthuysen transformation, we attain (non-)relativistic interactions related to the non-minimal coupling in three-dimensional spacetime, for both the bosonic and fermionic fields. Next, we establish some comparisons and analyse particular situations in which the external electromagnetic fields are described either by Maxwell or Maxwell-Chern-Simons Electrodynamics. In addition, we consider the situation of a non-minimal coupling for the fermionic field in four dimensions, carry out its dimensional reduction to three dimensions and show that the three-dimensional scenario previously worked out can be recovered as a particular case. Finally, we discuss a number of structural aspects of both procedures.

\end{abstract}


\maketitle

\section{Introduction} \label{Sec_Intro}
\indent

The works by Dirac and Wigner, in the development of Quantum Mechanics, pushed forward the concept and the far-reaching importance of symmetry. Ever since, symmetry principles have been widely applied in Physics. This legacy allowed us to progress in the description of the fundamental interactions and  simplify a series of  problems. In addition, our understanding of spacetime and its symmetries also plays an important role. Indeed, when adopting a particular spacetime or dimensionality, we may unveil interesting features such as topological quantities, possible boundary effects, exotic field representations and implementation of new symmetries.

In particular, low-dimensional systems have attracted a great deal of attention due to the discovery of new materials, parallel investigations of planar effects and collective modes in Condensed Matter Physics. However, two distinct viewpoints have been considered. Instead of analysing the system in low dimensions and exploring the aforementioned features, we could start with a higher-dimensional scenario and carry out a dimensional reduction. In this case, other degrees of freedom may appear in the reduced theory which may be responsible 
for new effects. We also highlight that many dimensional reduction procedures are available in the literature and there is no general proof that different prescriptions lead to the same result.

Our approach to these points is based on the Foldy-Wouthuysen (FW) transformation \cite{FW_PR_1950,Foldy_PR_1952,Vries_FP}, which enable us to diagonalize the Hamiltonian in an approximate way or, in some cases, in an exact form (see, for instance, \cite{Case_PR,Eriksen_PR,Nikitin_JPA,Silenko_JMP,Silenko_PRD_2014}). 
Even though the main purpose of the initial works was the investigation of the electromagnetic interaction, the method can be applied in other situations. For example, it has been a useful tool to  describe particles with different spin configurations \cite{Silenko_PRD_2014,Silenko_TMP,Silenko_PRD_2013,Silenko_PRD_2013_2,Silenko_PRD_2018,Kapusta_PRD}, as well as to analyse fundamental and excited states of atomic nuclei \cite{Guo_PRC}, some interactions with gravitational and torsion fields \cite{Ryder_Shapiro,Obukhov_PRL,Accioly,Silenko_Teryaev_grav,Obukhov_PRDs}. Also, it was considered in scenarios with Lorentz symmetry violation \cite{Belich_LV_2015,Goncalves_2009,Goncalves_2014_2019,Xiao_PRD_2018}, accelerated frames \cite{Hehl_Ni_1990,Chowdhury_Basu_2013_2014}, topological defects \cite{Wang_PRA_87,Chowdhury_Basu_2014} and  topological insulators \cite{Rothe_NJP_2010,Dayi_Ann_2012}. We emphasize that there are different approaches to the FW transformation.  In this work, we adopt the non-relativistic method, which shall be explained in the next Sections. However, it is worthy to mention the existence of relativistic FW transformations. For more details on relativistic methods and their validity, see \cite{Silenko_PRA_2015} and references therein.

Returning to the initial motivation to consider low-dimensional systems, an interesting aspect is the fact that the FW transformation can also be applied in other dimensionalities rather than four dimensions. In a seminal paper by Binegar \cite{Binegar}, where the  irreducible representations of the Poincaré group in three-dimensional spacetime were carefully built up, the author also contemplated the correspondent FW transformation for a free particle. Here, we pursue some investigations of the electromagnetic interaction in three-dimensional systems. Particularly, we focus on the effects of non-minimal electromagnetic couplings. The inclusion of these couplings contributes to a series of effects such as the  Aharonov-Casher phase \cite{Aharonov_Casher}, quantum Hall effect \cite{Yoshioka,Ezawa} and high-temperature superconductivity \cite{Mavromatos_Momen}. Moreover, the Chern-Simons term and non-minimal couplings in three dimensions allow 
spinless particles to acquire  anomalous magnetic moments \cite{Kogan_1991,Stern_1991}. With these effects and other applications, which shall be discussed in more details in the next Sections, we believe that studying the FW transformation with non-minimal electromagnetic couplings could reveal new insights of the relativistic corrections and bring some new contribution to the literature of the subject.

This work is organized with the following outline: in Section \eqref{FW_3D}, we discuss the FW transformation in three-dimensional spacetime for bosonic (scalar) and fermionic systems coupled non-minimally to an external electromagnetic field.
Here, we recover the well-known non-relativistic corrections and present the relativistic contributions. After that, in Section \eqref{Dim_reduction}, we investigate a similar case in four-dimensional spacetime, where we develop the FW transformation for a fermionic system with other non-minimal electromagnetic coupling and carry out its dimensional reduction. Then, we discuss whether these descriptions and non-minimal couplings are equivalent or may be connected as a particular case. Finally, in Section \eqref{Sec_Concluding}, we present our concluding comments and perspectives.

\section{Low-dimensional systems} \label{FW_3D}
\indent

In the study of charged matter fields and 
their interactions with an external electromagnetic field, 
the introduction of gauge potential $ A^\mu = ( \phi, \vec{A} \, ) $ allows us to implement the so-called minimal prescription, 
in which $ \partial_\mu $ is replaced by the (minimal) covariant derivative
\begin{equation}
D_{\mu} = \partial_\mu + iq A_\mu \, ,
\label{cov_der} \end{equation}
where $q$ denotes the electrical charge and we 
adopt natural units $\hbar = c =1$.

However, a particular feature appears in three-dimensional spacetime, namely, it is well-known that the dual field strength of $ F_{\mu \nu} = \partial_\mu A_\nu - \partial_\nu A_\mu $ corresponds to the vector $ \widetilde{F}_\mu \equiv \frac{1}{2} \epsilon_{\mu \nu \kappa} F^{\nu \kappa}  $, 
where  
$\epsilon^{\mu \nu \kappa}$ represents the Levi-Civita symbol. Throughout this work, we shall use the metric $\eta_{\mu \nu} = \textrm{diag}(+,-,-)$ and $\epsilon^{012} = \epsilon_{012} = +1$. Thus, the minimal covariant derivative \eqref{cov_der} can be generalized, assuming the form
\begin{equation}
\mathcal{\mathfrak{D}}_{\mu} = \partial_\mu + iq A_\mu + ig \widetilde{F}_\mu \, ,
\label{ext_cov_der} \end{equation}
where the coupling constant $g$ will be interpreted as an anomalous magnetic dipole moment.

At this point, it is convenient to introduce the dual spatial vector, $\widetilde{\vec{v}}_i = \epsilon_{ij} \vec{v}_j$, with $\epsilon_{ij} \equiv \epsilon_{0ij}$ and  Latin indexes denoting  the purely spatial sector $i,j = (1,2)$. Consequently, $\widetilde{\vec{v}}$ is perpendicular to $\vec{v}$. Bearing in mind the definitions of the electric and magnetic fields, $\vec{E} = - \vec{\nabla} \phi - \partial_t \vec{A}$ and $B = \vec{\nabla} \times \vec{A} \equiv \epsilon_{ij} \partial_i \vec{A}_j$, one can show that eq. \eqref{ext_cov_der} implements the following non-minimal electromagnetic prescription:
\begin{eqnarray}
i \partial_t & \, \rightarrow \, & \Pi^0 \equiv i \partial_t - q \phi + g B \, , \label{non_min_presc_1} \\
\vec{p} & \, \rightarrow \, & \vec{\Pi} \equiv \vec{p} - q \vec{A} + g \widetilde{\vec{E}} \, .
\label{non_min_presc_2} \end{eqnarray}

\noindent
Therefore, as expected, non-minimal contributions appear with electric and magnetic fields instead of gauge potentials $\phi$ and $\vec{A}$.

It is worthy to remark that this non-minimal coupling has been a subject of intense investigation in the literature, for instance, with applications in the description of Jain's composite fermion model and quantum Hall effect \cite{Paschoal_PLA_2003}. Also, some authors claim the possibility of describing anyons  without the Chern-Simons term \cite{Carrington_Kunstatter_1995,Nobre_Almeida_1999,Itzhaki_2003}. Moreover, its influence on the static inter-particle potentials was considered in both scalar and fermionic quantum electrodynamics with possible contributions at small distances (see  \cite{Dalmazi_2004,Dalmazi_2006} and references therein).

Let us also mention that, even though the extended covariant derivative \eqref{ext_cov_der} is defined in  three spacetime dimensions 
due to $\widetilde{F}_\mu$, this coupling can be seen as an inheritance of other non-minimal couplings in four dimensions. For example, in refs. \cite{Paschoal_PLA_2006,LV_model}, the authors discussed a Lorentz-symmetry violating coupling such that, for a particular background field, eq. \eqref{ext_cov_der} is recovered upon dimensional reduction. In this case, the parameter $g$ is related to a background field component.  At this stage, it is interesting to highlight that the Chern-Simons term  can be obtained through effective theories involving Lorentz symmetry violation in four dimensions, which may be generated by non-linear Electrodynamics in the presence of an external electromagnetic field or interactions with non-uniform distributions of matter fields. These provide realistic physical scenarios in four dimensions as starting points to describe some planar phenomena.
For more details, we indicate ref. \cite{Botta}. Similarly, in  scenarios involving non-commutative spacetime \cite{Ghosh_2005} or supersymmetry \cite{Christiansen_1999}, one can find other motivations to introduce the extended covariant derivative. Furthermore, we point out that the non-minimal coupling constant $g$ has a negative mass dimension, $[g]=-1/2$, which yields non-renormalizability and, consequently, we shall interpret them as effective field theories.

With the aforementioned motivations,  in Section \eqref{Dim_reduction}, we  
investigate a non-minimal electromagnetic coupling in four dimensions and its subsequent dimensional reduction. Before that, we first analyse the contributions of the non-minimal coupling \eqref{ext_cov_der}  in low-dimensional bosonic and fermionic systems.

\subsection{Scalar field} \label{FW_scalar}
\indent

Initially, let us consider the spin-0 matter field. The Klein-Gordon Lagrangian (density)  non-minimally coupled to an external electromagnetic field, is described by
\begin{eqnarray}
\mathcal{L}_{KG} = \left( \mathcal{\mathfrak{D}}_{\mu} \varphi \right)^\ast \, \mathcal{\mathfrak{D}}^{\mu} \varphi - m^2 \varphi^\ast \varphi \, .
\label{L_KG} \end{eqnarray}

\noindent
This Lagrangian is written in terms of minimal covariant derivative \eqref{cov_der} as follows
\begin{eqnarray}
\mathcal{L}_{KG} &=& \left( D_{\mu} \varphi \right)^\ast  D^{\mu} \varphi - m^2 \varphi^\ast \varphi  
- g \, J^\mu \widetilde{F}_\mu + g^2 \varphi^\ast \varphi \widetilde{F}_\mu^2 \, ,
\label{L_KG_1} \end{eqnarray}
where we define the  three-current
\begin{equation}
J^\mu = i \left[ \left( D^{\mu} \varphi \right) \varphi^\ast - \left( D^{\mu} \varphi \right)^\ast \varphi \right] \, .
\label{KG_current} \end{equation}

Here, we observe two contributions related to non-minimal coupling. The first one, $g \, J^\mu \widetilde{F}_\mu$, 
generates an anomalous magnetic dipole moment 
and similar coupling can also be introduced in other field representations with the correspondent  three-current, while  $g^2 \varphi^\ast \varphi \widetilde{F}_\mu^2$  is particular to spin-0 description. In this work, we only analyse the effects of an external electromagnetic field, therefore
we do not consider the quartic and sixth self-couplings, $ (\varphi^\ast \varphi)^2$ and $(\varphi^\ast \varphi)^3$.

The equation of motion reads as below
\begin{equation} ( \Pi^2 - m^2 ) \, \varphi = 0 \, , \label{KG_eq} \end{equation}
with $ \Pi^2 \equiv \Pi_\mu \Pi^\mu $ and $ \Pi^\mu = (\Pi^0 ; \vec{\Pi}) $.

In order to analyse the (non-)relativistic contributions, we introduce a new field $\chi_\mu$
and note that eq. \eqref{KG_eq} is equivalent to the following two first-order equations:
\begin{eqnarray} m \chi_\mu &=& \Pi_\mu \varphi \, , \label{KG_eq_2} \\
 \Pi^\mu \chi_\mu &=& m \varphi \, . \label{KG_eq_3} \end{eqnarray}
From eq. \eqref{KG_eq_2} with $\mu =i$, we get the constraints
$\chi_i = \frac{1}{m} \Pi_i \varphi$. Thus, only $\chi_0$ and $\varphi$ are dynamical fields.

Manipulating these equations, we arrive at
\begin{eqnarray} i \partial_t \varphi &=& m \chi_0 + (q \phi - gB) \varphi  \, , \label{KG_eq_5} \\
i \partial_t \chi_0 &=& m \varphi + (q \phi - gB) \chi_0 + \frac{\vec{\Pi}^2}{m}  \varphi  \, , 
\label{KG_eq_6} \end{eqnarray}
which is recast in a Schrödinger-like equation, $ i \partial_t \rho = H_0 \rho $, for a two-component field
\begin{equation}  
\rho = \left( \begin{array}{c} \rho_a \\ \rho_b \end{array} \right) = \frac{1}{2} 
\left( \begin{array}{c} \varphi + \chi_0 \\ \varphi - \chi_0 \end{array} \right) \, .
\end{equation}
In this case, we obtain the Hamiltonian 
\begin{equation}
H_0 = \frac{\vec{\Pi}^2}{2m} \left( \mathcal{R} + \mathcal{N} \right) + m \mathcal{N} + (q \phi - gB) \mathbb{I}
\, , \label{H_KG} \end{equation}
where $\mathbb{I}$ denotes the identity $2 \times 2$, $\mathcal{R} = i \sigma_y$ and $\mathcal{N} = \sigma_z$, with $\sigma_y$ and $\sigma_z$ being the well-known Pauli matrices.

Similarly to the procedure described in ref. \cite{Schwabl_book} for four-dimensional spacetime, one can decompose 
the Hamiltonian as
\begin{equation}
H_0 =  m \mathcal{N} + \mathcal{E} + \mathcal{O} \, , 
\end{equation}
with the operators $\mathcal{E}$ and $\mathcal{O}$ satisfying $\mathcal{E} \mathcal{N} = \mathcal{N} \mathcal{E}$ and 
$\mathcal{O} \mathcal{N} = - \mathcal{N} \mathcal{O}$. Using successive unitary transformations, $H'_0 = e^{iS} (H_{0} - i \partial_t) e^{-iS} $ with the Hermitian operator $S = -i \mathcal{N} \mathcal{O}/2m$, it is possible to obtain the leading contributions of the FW transformation for spin-0 matter field:  
\begin{eqnarray}
H_{FW,0} & \approx & m \mathcal{N}  +  \mathcal{E} + \mathcal{N} \frac{\mathcal{O}^2}{2m} 
- \frac{1}{8m^2} \left[  \mathcal{O} , \, \left[ \mathcal{O} ,  \mathcal{E} \right] + i \dot{\mathcal{O}}   \right]
\, , \label{FW_spin_0} \end{eqnarray}
which also holds in three-dimensional spacetime. For the Hamiltonian \eqref{H_KG}, we identify
\begin{equation}
\mathcal{O} = \frac{\vec{\Pi}^2}{2m}  \mathcal{R} \, \; , \, \; \mathcal{E} = \frac{\vec{\Pi}^2}{2m}  \mathcal{N}  + (q \phi - gB) \mathbb{I} \, . \label{KG_O_E}\end{equation}

Therefore, for positive energy solution, the diagonalized Hamiltonian up to order $O(1/m^3)$ is given by  
\begin{eqnarray}
H_{FW,0}  &\approx& m   + \frac{ ( \vec{p} - q \vec{A} + g \widetilde{\vec{E}} \, )^2}{2m} + q \phi 
-  gB - \frac{ ( \vec{p} - q \vec{A} + g \widetilde{\vec{E}} \, )^4}{8m^3} \, .
\label{FW_0} \end{eqnarray}

This Hamiltonian contains the usual energy potential $q \phi$ and non-relativistic kinetic term
with the canonical momentum $ \vec{\Pi} = \vec{p} - q \vec{A}  + g \widetilde{\vec{E}} $, which 
can generate the Aharonov-Bohm  and Aharonov-Casher phases \cite{Aharonov_Casher,Aharonov_Bohm}. Moreover, we get an  anomalous magnetic dipole interaction $g B$. The last term in eq. \eqref{FW_0} contributes to the relativistic mass correction. In a heuristic way, 
is possible to obtain it from the Taylor expansion of the relativistic kinetic energy $\sqrt{\vec{\Pi}^2 + m^2}$. Finally, we highlight that there are no terms proportional to $O(1/m^2)$, which is a 
scalar case particularity. We shall return to this point in the next subsection, where we discuss the fermionic matter field.

\subsection{Fermionic  field} \label{FW_fermion}
\indent

We now consider the Dirac Lagrangian with non-minimal coupling \eqref{ext_cov_der}, namely,
\begin{equation}
\mathcal{L}_D = i \bar{\psi} \gamma^\mu \mathcal{\mathfrak{D}}_{\mu} \psi - m \bar{\psi} \psi \, ,
\label{L_Dirac_3D} \end{equation}
or, equivalently, in terms of the minimal covariant derivative \eqref{cov_der},
\begin{equation}
\mathcal{L}_D = i \bar{\psi} \gamma^\mu D_{\mu} \psi - m \bar{\psi} \psi - g \, j^\mu \widetilde{F}_\mu
\, , \label{L_Dirac_3D_1} \end{equation}
where $ j^\mu = \bar{\psi} \gamma^\mu \psi $. As already anticipated, we note the presence of non-minimal interaction $g \, j^\mu \widetilde{F}_\mu$ similar to the spin-0 case, 
see eq. \eqref{L_KG_1}, which shall be responsible for an anomalous magnetic moment interaction.

To proceed further, we remember that the FW transformation of fermionic field is usually carried out with gamma-matrices in the Dirac representation. For three-dimensional spacetime, we have $\gamma^0 = \sigma_z$, $\gamma^1 = i \sigma_x$ and $\gamma^2 = i \sigma_y$. Besides the Clifford algebra, $\left\{ \gamma^\mu , \gamma^\nu \right\} = 2 \eta^{\mu \nu} \mathbb{I} $, these matrices also satisfy the identity 
\begin{equation} \gamma^\mu \gamma^\nu = \eta^{\mu \nu} \mathbb{I} - i \epsilon^{\mu \nu \kappa} \gamma_\kappa \, . \label{gamma_identity} \end{equation}

Making use of the definitions $\beta \equiv \gamma^0$ and $\vec{\alpha} \equiv \beta \vec{\gamma} $, we can rewrite the Dirac equation
\begin{equation}
\left( i \gamma^\mu \mathcal{\mathfrak{D}}_{\mu} - m \right) \psi = 0 \, 
\label{Dirac_eq_3D} \end{equation}
into the form $ i \partial_t \psi = H_D \, \psi $, with the Hamiltonian 
\begin{equation}
H_D = m \beta  + \vec{\alpha} \cdot \vec{\Pi} + \left( q \phi - g B \right) \mathbb{I} \, .
\label{H_Dirac_3D} \end{equation}

Next, we consider a similar procedure to what was carried out in the previous subsection. Thus, we decompose 
\begin{equation}
H_D = m \beta  + \mathcal{E} + \mathcal{O} \, ,
\label{H_Dirac_FW} \end{equation}
where the operators above satisfy $\mathcal{E} \beta =  \beta \mathcal{E} $ and 
$\mathcal{O} \beta = - \beta \mathcal{O}$. Then, by means of unitary transformations, we end up with the diagonalized Hamiltonian \cite{Schwabl_book}:

\begin{eqnarray}
H_{FW,D} & \approx &  m \beta  +  \mathcal{E} + \beta \frac{\mathcal{O}^2}{2m} 
- \frac{1}{8m^2} \left[  \mathcal{O} , \, \left[ \mathcal{O} ,  \mathcal{E} \right] + i \dot{\mathcal{O}}   \right]
\, . \label{FW_spin_meio} \end{eqnarray}
This result also holds true in other dimensionalities. Furthermore, in three-dimensional spacetime, it exhibits the same functional form as the spin-0 case, eq. \eqref{FW_spin_0}, because $\mathcal{N} = \beta = \sigma_z$. However, the operators $\mathcal{O}$ and $\mathcal{E}$ are different. For the fermionic situation, we have
\begin{equation}
\mathcal{O} = \vec{\alpha} \cdot \vec{\Pi} \, \; , \, \; \mathcal{E} = \left( q \phi - g B \right) \mathbb{I} 
\, . \label{D_O_E_3D} \end{equation}

Using these operators in eq. \eqref{FW_spin_meio} and assuming positive-energy solution, we obtain
\begin{eqnarray}
H_{FW,D} &\approx & m + \frac{ ( \vec{p} - q \vec{A} + g \widetilde{\vec{E}} \, )^2}{2m} + q \phi - g B  
- \frac{q }{2m} B - \frac{g}{2m} \vec{\nabla} \cdot \vec{E} \nonumber \\
&-& \frac{iq}{8m^2} \, \vec{\nabla} \times \vec{\mathbb{E}}  \,
- \frac{q}{4m^2} \, \vec{\mathbb{E}} \times ( \, \vec{p} - q \vec{A}  
+ g \widetilde{\vec{E}} \,) \,
- \frac{q}{8m^2} \, \vec{\nabla} \cdot \vec{\mathbb{E}} \, ,
\label{FW_D} \end{eqnarray}
where we define
\begin{equation}
\vec{\mathbb{E}} \equiv \vec{E} + \frac{g}{q} ( \, \vec{\nabla} B + \partial_t \widetilde{\vec{E}} \, ) \, .
\label{E_eff} \end{equation}

At this stage, it is worthy to compare the Hamiltonians \eqref{FW_0} and \eqref{FW_D}. In the non-relativistic limit, we have two additional contributions for the fermionic case, given by $q B/2m$ and $g \vec{\nabla} \cdot \vec{E}/2m$. The first corresponds to the usual magnetic moment interaction.  We emphasize that the spin projection $S_z = \sigma_z/2$ is not explicit.  The second is a non-minimal contribution similar to Darwin term \cite{Schwabl_book}. 
The next-order (relativistic) contributions in eq. \eqref{FW_D} exhibit terms proportional to $O(1/m^2)$, which are not present in the scalar case. For $g=0$  $(\vec{\mathbb{E}} \rightarrow \vec{E})$, we obtain interactions analogous to the four-dimensional result, including the well-known spin-orbit coupling and Darwin term (see, for instance, ref. \cite{Itzykson_book}). In the next section, we shall return to this point,  where some comparisons will be established by means of projected state. Moreover, we highlight that the relativistic contribution $- ( \vec{p} - q \vec{A} + g \widetilde{\vec{E}} \, )^4/8m^3$ also appears in the fermionic case when including higher-order corrections to eq. \eqref{FW_spin_meio}, namely, it comes from the next contribution $- \beta \mathcal{O}^4/8m^3$.

To get more insights on non-minimal interactions, it is advisable to specify the particular Electrodynamics related to 
the external field. Let us consider the Maxwell-Chern-Simons theory, described by the following  Lagrangian
\begin{eqnarray}
\mathcal{L}_{MCS} = - \frac{1}{4} F_{\mu \nu}^2 + \frac{\lambda}{2} \epsilon^{\mu \alpha \beta} A_\mu \partial_\alpha A_\beta
- J^\mu A_\mu \, ,\label{L_MCS} \end{eqnarray}
where $J^\mu = (\rho, \vec{J})$  and $\lambda$ denotes the Chern-Simons parameter.  

The equations of motion are given by
\begin{eqnarray}
\vec{\nabla} \times \vec{E} = - \partial_t B \, , \label{MCS_1}  \\
\vec{\nabla} \cdot \vec{E}  - \lambda B = \rho \, , \label{MCS_2}  \\
\widetilde{\vec{\nabla}} B - \lambda \widetilde{\vec{E}} = \vec{J} + \partial_t \vec{E}
\, . \label{MCS_3}  
\end{eqnarray}

\noindent
From eq. \eqref{MCS_3} we can rewritten eq. \eqref{E_eff} as
\begin{equation}
\vec{\mathbb{E}} = \left( 1 + \frac{\lambda g}{q} \right) \vec{E} - \frac{g}{q} \, \widetilde{\vec{J}}  \, .
\label{E_eff_2} \end{equation}

For the particular case of Maxwell theory $(\lambda = 0)$ in vacuum, we have $\vec{\mathbb{E}} \rightarrow \vec{E}$  and most of the relativistic effects associated with non-minimal contributions disappear in eq. \eqref{FW_D}, except for an electric density energy $qg \vec{E}^2/4m^2$. Then, within our approximations, non-minimal relativistic interactions are relevant only in matter (with $\vec{J} \neq \vec{0}$). This result is very similar to what happens with the (Zeldovich) anapole moment in four-dimensional spacetime  \cite{Nowakowski}.

On the other hand, for the Maxwell-Chern-Simons  theory in vacuum, we obtain relativistic interactions with an effective electric field $\vec{\mathbb{E}} \rightarrow \vec{E}_{\textrm{eff}} = \left( 1 + \frac{\lambda g}{q} \right) \vec{E}$. It should be noted that a similar redefinition can be used in the magnetic moment interaction in terms of an effective magnetic field 
$B_{\textrm{eff}} = \left( 1 + \frac{\lambda g}{q} \right) B$, after using the equation $\vec{\nabla} \cdot \vec{E}  = \lambda B $ to recast some non-relativistic contributions. Therefore, for the critical value $ g_c = - q/\lambda$, we have some cancellations such that fermionic and scalar fields exhibit the same Hamiltonian up to order $O(1/m^2)$, namely, 
\begin{equation}
H_c  \approx  m + \frac{ ( \vec{p} - q \vec{A} + g_c \widetilde{\vec{E}} \, )^2}{2m} + q \phi - g_c B \, .
\label{H_c} \end{equation}
This critical value also plays an important role in other contexts. For instance, it corresponds to the situation in which all the one-loop quantum corrections to the photon mass disappear \cite{Georgelin}. Furthermore, it also appears as a special condition to obtain  vortex solutions \cite{Torres}.


\section{Dimensional Reduction Analysis} \label{Dim_reduction}
\indent

Having established the low-dimensional systems, we draw our attention to investigate a non-minimal coupling in four spacetime dimensions. Our motivation is based on a line of works in the literature in which a higher-dimensional scenario plays an important role in the description of the reduced system. For instance, we mention planar effects induced by Lorentz symmetry breaking in four spacetime dimensions \cite{Paschoal_PLA_2006,LV_model,Botta}, the theoretical proposal of the topological superconductor \cite{Witten_2013}; a possible dark sector from five-dimensional Electrodynamics coupled to 3-form gauge field \cite{Cocuroci_EPJC_2015}; new interactions and spin-dependent contributions introduced through dimensional reduction \cite{Silenko_compactification,Leo_Helayel_PRD} and effects of boundary conditions on reduced field theories \cite{erich}. Within this point of view, we consider a fermionic field non-minimally coupled to an external electromagnetic field. Initially, we apply the FW transformation to obtain the correspondent (non-)relativistic interactions. After that, we figure out the dimensional reduction to three-dimensional spacetime and establish some connections with the previous subsection. Here, we adopt the indexes $\hat{\mu} , \hat{\nu} = (0,1,2,3)$ and metric $ \eta_{\hat{\mu} \hat{\nu}} = \textrm{diag}(+,-,-,-) $ for four-dimensional spacetime. In order to avoid confusion with the previous section, we  use bold characters to denote the three-dimensional space vectors. For example, the electromagnetic potential is given by $ A^{\hat{\mu}} = (\phi, \vec{\bm{A}}) $.

Besides the usual Dirac Lagrangian minimally coupled to an external electromagnetic field, we also include the so-called Pauli interaction such that 
\begin{equation}
\mathcal{L}_{DP} = i \bar{\psi} \gamma^{\hat{\mu}} D_{\hat{\mu}} \psi - m \bar{\psi} \psi + \frac{f}{2} \, \bar{\psi} \sigma^{\hat{\mu} \hat{\nu}} \psi \, F_{\hat{\mu} \hat{\nu}} \,  
\label{L_DP_4D} \end{equation}
with $f$ being the non-minimal coupling constant, 
$ D_{\hat{\mu}} = \partial_{\hat{\mu}} + i q A_{\hat{\mu}} $ and $ \sigma^{\hat{\mu} \hat{\nu} } = \frac{i}{2} \left[ \gamma^{\hat{\mu}}, \gamma^{\hat{\nu}} \right] $.

Before going ahead, let us present some comments. Again, the non-minimal coupling constant has a negative mass dimension, $[f] =-1$, thus the  Dirac-Pauli Lagrangian above
should be considered as an effective field theory. We also emphasize that this non-minimal interaction was recently investigated in the context of  spin Hall effect
\cite{Turcati_et_al}. In addition, 
returning to low-dimensional systems, which is our initial motivation, from identity \eqref{gamma_identity} one can show that $ j^\mu \widetilde{F}_\mu = \frac{1}{2} \, \bar{\psi} \sigma^{\mu \nu} \psi \, F_{\mu \nu} $. Therefore, the non-minimal interaction in three dimensions is analogous to the Pauli term. For this reason, we expect to connect these non-minimal couplings through dimensional reduction. Indeed, in ref. \cite{Carrington_Kunstatter_1995}, the authors established some comparisons in the non-relativistic limit. Here, we shall extend this result by including the relativistic contributions.

We are now in 
position to discuss the 
FW transformation involving this non-minimal interaction. From the Lagrangian \eqref{L_DP_4D}, we get the equation of motion

\begin{equation}
\left( i \gamma^{\hat{\mu}} D_{\hat{\mu}} - m + \frac{f}{2} \,  \sigma^{\hat{\mu} \hat{\nu} } F_{\hat{\mu} \hat{\nu}}\right) \psi = 0
\label{Dirac_eq_4D} \end{equation}
and using the gamma matrices in the Dirac representation,
\begin{equation} \gamma^0 = \left( \begin{array}{cc} \mathbb{I}  &  0 \\ 0  &  - \mathbb{I} \end{array} \right) 
\, \; , \, \;  \gamma^i = \left( \begin{array}{cc} 0  &  \sigma_i \\   - \sigma_i & 0 \end{array} \right)
\, , \label{gamma_4D} \end{equation}
we obtain the Dirac-Pauli Hamiltonian

\begin{eqnarray}
H_{DP} &=& \beta m + \vec{\bm{\alpha}} \cdot \vec{\bm{\pi}} + q \phi \, \bm{I} 
- i f \, \beta \, \vec{\bm{\alpha}} \cdot \vec{\bm{E}} + f \, \beta 
\, \vec{\bm{\Sigma}} \cdot \vec{\bm{B}} \, ,
\label{H_Dirac_4D} \end{eqnarray}
where $\bm{I}$ denotes the identity $4\times 4$, $\vec{\bm{\pi}} \equiv \vec{\bm{p}} - q \vec{\bm{A}}$ and $\vec{\bm{\Sigma}}$ corresponds to the spin matrix 
\begin{equation} \vec{\bm{\Sigma}} = \left( \begin{array}{cc} \vec{\sigma}  &  0 \\  0 & \vec{\sigma}   \end{array} \right) 
\, . \label{spin_matrix} \end{equation}

\noindent
In this case, we identify
\begin{eqnarray}
\mathcal{O} &=& \vec{\bm{\alpha}} \cdot \vec{\bm{\pi}}  
- i f \, \beta \, \vec{\bm{\alpha}} \cdot \vec{\bm{E}} \, \; \,  , \, \; \,
\mathcal{E} = q \phi \, \bm{I} + f \, \beta \, \vec{\bm{\Sigma}} \cdot \vec{\bm{B}}
\, . \label{DP_O_E} \end{eqnarray}

Then, replacing these operators in eq. \eqref{FW_spin_meio} with the correspondent gamma matrices and assuming positive energy solution, it is possible to show that

\begin{eqnarray}
H_{FW,DP} & \approx & m \, \mathbb{I} + \frac{1}{2m} \, \left( \vec{\bm{\pi}} \, \mathbb{I} - f \vec{\bm{E}} \times \vec{\sigma} \right)^2 + q \phi \, \mathbb{I} + \left( f - \frac{q}{2m} \right) \vec{\sigma} \cdot \vec{\bm{B}} - \frac{f^2}{2m} \vec{\bm{E}}^2 \, \mathbb{I}
\nonumber \\
&+& \left( \frac{f}{2m} - \frac{q}{8 m^2} \right) \vec{\nabla} \cdot \vec{\bm{E}} \, \mathbb{I} 
- i \, \frac{q}{8 m^2} \, \vec{\sigma} \cdot \left( \vec{\nabla} \times \vec{\bm{E}} \right) 
-  \, \frac{q}{4 m^2} \, \vec{\sigma} \cdot \left( \vec{\bm{E}} \times \vec{\bm{\pi}} \right) 
\nonumber \\
&-& \frac{q f}{4 m^2} \, \vec{\bm{E}}^2 \, \mathbb{I} - i \, \frac{f}{8 m^2} \, \left[ \vec{\nabla} \cdot \left( \partial_t \vec{\bm{E}} \right) \right] \, \mathbb{I} + \frac{f}{8 m^2} \, \vec{\sigma} \cdot \left( \vec{\nabla} \times \partial_t \vec{\bm{E}} \right) \nonumber \\
&+& \frac{ f}{8 m^2} \, \left[ \nabla^2 \left( \vec{\sigma} \cdot  \vec{\bm{B}} \right) \right] +
\frac{ f}{4 m^2} \, \mathbb{I} \left[ \partial_t \vec{\bm{E}} - \vec{\nabla} \times  \vec{\bm{B}} \right] \cdot \vec{\bm{\pi}} + i \, \frac{ f}{4 m^2} \, \left( \vec{\nabla} \cdot \vec{\bm{B}} \right) \vec{\sigma} \cdot \vec{\bm{\pi}} 
\nonumber \\
&+& i \, \frac{ f}{4 m^2} \, \left[ \left( \vec{\sigma} \cdot \vec{\nabla} \right) \vec{\bm{B}} \right] \cdot \vec{\bm{\pi}} 
- \frac{ f}{2 m^2} \, \vec{\bm{B}} \cdot \Bigl[  \Bigl( \vec{\sigma} \cdot \vec{\bm{\pi}}  \Bigr) \vec{\bm{\pi}}  \Bigr]
+ \frac{ f^2}{4 m^2} \, \vec{\sigma} \cdot \left( \vec{\bm{E}} \times \partial_t \vec{\bm{E}} \right) 
\nonumber \\
&-& \frac{ f^2}{4 m^2} \, \left[ \vec{\sigma} \cdot \vec{\nabla} \Bigl( \vec{\bm{E}} \cdot \vec{\bm{B}} \Bigr)  \right]
+  \frac{ f^2}{4 m^2} \, \left[ \vec{\bm{E}} \cdot \vec{\nabla} \Bigl( \vec{\sigma} \cdot \vec{\bm{B}} \Bigr)  \right]
- \frac{ f^2}{4 m^2} \, \vec{\bm{E}} \cdot \left[ \Bigl( \vec{\sigma} \cdot \vec{\nabla} \Bigr) \vec{\bm{B}} \right]
\nonumber \\
&-& \frac{ f^2}{4 m^2} \, \left[ \vec{\bm{B}} \cdot \vec{\nabla} \Bigl( \vec{\sigma} \cdot \vec{\bm{E}} \Bigr)  \right] 
- \frac{ f^3}{2 m^2} \, \Bigl( \vec{\bm{E}} \cdot \vec{\bm{B}} \Bigr) \Bigl( \vec{\sigma} \cdot \vec{\bm{E}} \Bigr)
\, . \label{FW_DP} \end{eqnarray}

This Hamiltonian contains 
new interactions and features, which deserve some comments. First of all, we recover the well-known result when $f=0$ \cite{Schwabl_book}, containing the traditional spin-dependent interactions $(\vec{S} = \vec{\sigma}/2)$  related to   $\vec{\sigma} \cdot \vec{\bm{B}}$, $\, \vec{\sigma} \cdot (\vec{\nabla} \times \vec{\bm{E}})$ and $\, \vec{\sigma} \cdot ( \vec{\bm{E}} \times \vec{\bm{\pi}})$. As expected, the non-minimal coupling introduces an  anomalous magnetic moment interaction and correction to the Darwin term, given by $ \, f  \vec{\sigma} \cdot \vec{\bm{B}}$ and  $\, \frac{f}{2m} \vec{\nabla} \cdot \vec{\bm{E}}$, respectively. It should be noticed that we do not specify the Electrodynamics. For this reason, we maintain the contributions with  $ \, \vec{\nabla} \cdot \vec{\bm{B}} $ and $\, \partial_t \vec{\bm{E}} - \vec{\nabla} \times  \vec{\bm{B}} \,$ in the fourth line of eq. \eqref{FW_DP}. By considering the Maxwell theory in vacuum, these contributions vanish. However, the term with $ \vec{\nabla} \cdot \vec{\bm{B}} $ may be relevant in the presence of magnetic monopoles and trigger investigations in Condensed Matter, such as the  spin ice configurations \cite{spin_ice}. In addition, 
some extensions of Maxwell Electrodynamics involving, for example, non-linear corrections \cite{EH,BI}, Lorentz symmetry violation \cite{modelo_CFJ} and quantum gravity effects \cite{Gambini_Pullin,Alfaro_et_al}, can introduce non-trivial 
contributions of $\partial_t \vec{\bm{E}} - \vec{\nabla} \times  \vec{\bm{B}}$, even in vacuum. 
Interesting enough, some interactions proportional to $f^2/m^2$ and $f^3/m^2$ exhibit new spin-dependent couplings between $\vec{\bm{E}}$ and $\vec{\bm{B}}$ such that parity symmetry is preserved. Finally, it is worthy to comment that the relativistic Dirac-Pauli Hamiltonian in the FW representation was previously carried out in ref. \cite{Silenko_JMP}. Moreover, it was also investigated in ref. \cite{Chen_PRA}, where the authors considered high-order contributions in the regime of weak, homogeneous and static fields. Within these assumptions, we recognize some interactions similar to our result.


At this stage, we are ready to implement a dimensional reduction in eq. \eqref{FW_DP}. We adopt the simplest procedure, similar to the Scherk-Schwarz reduction \cite{Scherk-Schwarz}, described by the following ansatz: $ \partial_z ( \textrm{any field}) = 0 $, i.e., the third spatial dimension is freezed-out while others dimensions remain unchanged. With this assumption in the Hamiltonian (or Lagrangian) description, the integration over the z-coordinate (normally taken to be compact) yields a length dimension factor, which is suitably 
absorbed by the coupling constants and fields, leading to the correct mass dimensions in a three-dimensional spacetime. Bearing this in mind, we decompose  $ \vec{\bm{A}} = ( \vec{A}, \varphi)$, where $\vec{A}$ denotes the planar vector potential and $\varphi$ corresponds to a new scalar field. For our purposes, we take the trivial solution $\varphi = 0$, which leads to the reduction $ \vec{\bm{B}} \rightarrow (0,0,B_z) $, $\vec{\bm{E}} \rightarrow (\vec{E},0)$ and $\vec{\bm{\pi}}_z \rightarrow 0$. After some algebraic manipulations, we find the following reduced Hamiltonian

\begin{eqnarray}
H_{\textrm{red}} & \approx & m \, \mathbb{I} + \frac{1}{2m} \, \left[ \left( \vec{p} - q \vec{A} \right) \, \mathbb{I} - f \, \widetilde{\vec{E}} \,\sigma_z  \right]^2 + q \phi \, \mathbb{I} + \left( f - \frac{q}{2m} \right) \, B_z \, \sigma_z + \frac{f}{2m} \, \vec{\nabla} \cdot \vec{E} \, \mathbb{I}
\nonumber \\
&-& i \, \frac{q}{8m^2} \, \sigma_z \, \vec{\nabla} \times \left[ \vec{E} \, \mathbb{I} - \frac{f}{q} \, \partial_t  \widetilde{\vec{E}}  \, \sigma_z \right] 
-  \frac{q}{8m^2} \, \vec{\nabla} \cdot \left[ \vec{E} \, \mathbb{I} - \frac{f}{q} \, \left( \vec{\nabla} B_z + \partial_t \widetilde{\vec{E}} \right) \, \sigma_z \right] 
\nonumber \\
&-& \frac{q}{4m^2} \, \sigma_z \left[ \vec{E} \, \mathbb{I} - \frac{f}{q} \, \left( \vec{\nabla} B_z + \partial_t \widetilde{\vec{E}} \right) \, \sigma_z \right] \times \left[ \left( \vec{p} - q \vec{A} \right) \, \mathbb{I} - f \, \widetilde{\vec{E}} \,\sigma_z \right] \, ,
\label{H_red} \end{eqnarray}
where we recognized $\epsilon_{3 \, ij} = \epsilon_{ij}$ to recover the dual electric field $\widetilde{\vec{E}}_i = \epsilon_{ij} \vec{E}_j$.

The reduced Hamiltonian must be applied to a two-component spinor $\psi = \left( \begin{array}{c}
\psi_1 \\ \psi_2 \end{array} \right)$. By considering the projected state in the upper component $(\psi_2 = 0)$, it is not difficult to see that we arrive at the previous three-dimensional result, eq. \eqref{FW_D}, with the identifications $B_z \rightarrow B$ and $f \rightarrow - g$. Therefore, we have shown that the non-minimal interactions in eq. \eqref{FW_D}
are attainable through a particular dimensional reduction of eq. \eqref{FW_DP}.

It is important to discuss the structural aspects of both procedures. Initially, we emphasize that spin-dependent interactions (related to $S_z = \sigma_z/2$) are more evident in the reduced Hamiltonian \eqref{H_red}. The reason is that we have four-component spinor in four spacetime dimensions and, when adopting the Hamiltonian with positive-energy solution, we end up with an equation for a two-component spinor. Similarly, in three spacetime dimensions, we start off with a two-component spinor and project to the one-component subspace. This explains the lack of $\sigma_z$ in the Hamiltonian \eqref{FW_D}. 

Even though the Hamiltonians are connected by a dimensional reduction, we have different points of view to describe the external electromagnetic field. If we adopt Maxwell Electrodynamics, we obtain the same result for both procedures, i.e., maintaining the previous dimensional reduction, the descriptions in eqs. \eqref{FW_D} and \eqref{H_red} are equivalent.  However, by taking into account some extensions of Maxwell Electrodynamics, we may arrive at different results. For instance, we remember that Maxwell-Chern-Simons Electrodynamics, eq. \eqref{L_MCS}, is particular to three spacetime dimensions due to the Levi-Civita symbol $\epsilon^{\mu \alpha \beta}$. On the other hand, we could start from an extension of Maxwell Electrodynamics in four spacetime dimensions and then perform its dimensional reduction, which leads to non-trivial equations of motion in the reduced theory.


\section{Concluding Comments} \label{Sec_Concluding}
\noindent

We investigate the contributions of non-minimal electromagnetic interactions in low-dimensional systems through two different viewpoints, which consist of describing the system in low dimensions or taking into account a higher-dimensional scenario and its reduction. We use the FW transformation to obtain the (non-)relativistic interactions in the Hamiltonian formalism.

Our main results are summarized as follows. In the case of three-dimensional systems and disregarding the terms  proportional to $O(1/m^3)$, we show that non-minimal relativistic interactions for the fermionic field are relevant only in the presence of matter $(\vec{J} \neq \vec{0})$ with the external electromagnetic field described by Maxwell Electrodynamics. On the other hand, non-trivial terms appear 
for Maxwell-Chern-Simons Electrodynamics, even in vacuum, and we find a critical value, $g_c = - q/\lambda$, such that bosonic and fermionic systems converge to the same Hamiltonian. Furthermore, in the context of four spacetime dimensions, we consider a non-minimal coupling for the fermionic field and demonstrate that the low-dimensional Hamiltonian is obtained by a particular dimensional reduction. In principle, both procedures are not equivalent when considering extensions of Maxwell Electrodynamics.

To conclude, we would like to point out some perspectives. In this work, we have focused on the consequences and connections between non-minimal couplings considered in different dimensions. To follow this program, we have adopted a particular dimensional reduction of the Hamiltonian \eqref{FW_DP}. However, this Hamiltonian unveils several interactions that justify and deserve further investigations. For example, we could carry out alternative dimensional reduction procedures and contemplate effects of non-trivial degrees of freedom, such as $\varphi \neq 0$ and Kaluza-Klein-type modes that arise upon compactification of a space dimension. 
In addition, we also stress that the introduction of relativistic corrections in the magnetization dynamics equation has been the subject of intense research over the past few years (for instance, see \cite{Mondal_PRB_2016_2018,Mondal_PRB_2017} and references therein).  By selecting the magnetic sector of eq. \eqref{FW_DP}, we shall obtain the corresponding extension of the Landau-Lifshitz-Gilbert (LLG) equation and analyse if these non-minimal interactions could lead to new damping or torque contributions to magnetization.  This is a matter of our present interest and it remains to be investigated in a forthcoming work to appear elsewhere.

\section*{Acknowledgements}
\noindent

We are grateful to J.A. Helayël-Neto for reading the manuscript and pertinent comments. This work was supported by the { \it National Council for Scientific and Technological Development} (CNPq, Brazil). LPRO thanks the Institutional Qualification 
Program (PCI).


\end{document}